# Direct Frequency Comb Spectroscopy in the Extreme Ultraviolet

Arman Cingöz[1*], Dylan C. Yost[1*], Thomas K. Allison[1], Axel Ruehl[2,3], Martin E. Fermann[2], Ingmar Hartl[2] & Jun Ye[1]

Development of the optical frequency comb has revolutionised metrology and precision spectroscopy due to its ability to provide a precise and direct link between microwave and optical frequencies[1,2]. A novel application of frequency comb technology that leverages both the ultrashort duration of each laser pulse and the exquisite phase coherence of a train of pulses is the generation of frequency combs in the extreme ultraviolet (XUV) via high harmonic generation (HHG) in a femtosecond enhancement cavity[3,4]. Until now, this method has lacked sufficient average power for applications, which has also hampered efforts to observe phase coherence of the high-repetition rate pulse train produced in the extremely nonlinear HHG process. Hence, the existence of a frequency comb in the XUV has not been confirmed. We have overcome both challenges. Here, we present generation of >200 µW per harmonic reaching 50 nm (20 µW after harmonic separation), and the observation of single-photon spectroscopy signals for both an argon transition at 82 nm and a neon transition at 63 nm. The absolute frequency of the argon transition has been determined via direct frequency comb spectroscopy. The resolved 10-MHz linewidth of the transition, limited by the transverse temperature of the argon atoms, is unprecedented in this spectral region and places a stringent upper limit on the linewidth of individual comb teeth. Due to the lack of continuous wave lasers, these frequency combs are currently the only promising avenue towards extending ultrahigh precision spectroscopy to below the 100-nm spectral region with a wide range of applications that include spectroscopy of electronic transitions in molecules[5], experimental tests of bound state and many body quantum electrodynamics in $He^+$ and $He$[6,7], development of next-generation "nuclear" clocks[8,9,10], and searches for spatial and temporal variation of fundamental constants[11,12] using the enhanced sensitivity of highly charged ions[13,14].

Techniques developed to control a train of ultrashort pulses in the frequency domain have led to rapid advancements not only in ultrahigh precision metrology[1], but also in generation of attosecond pulses for time-resolved studies[15]. This symbiotic relationship between time and frequency techniques continues with the development of the XUV frequency combs where HHG, a standard technique for attosecond physics, is utilized to produce phase coherent XUV radiation. In conventional HHG, a single infrared pulse generates a burst of attosecond pulses separated by half cycles of the driving laser field, resulting in the odd harmonic spectrum shown in Fig 1. In contrast, in intracavity HHG, a phase-coherent infrared pulse train is used to produce a train of such bursts that repeat at the repetition frequency of the fundamental comb. This new temporal structure is responsible for the much finer frequency comb within each harmonic order. We anticipate that high precision characterization of the HHG process enabled by the XUV frequency comb will once again provide unique insights into attosecond physics.

[1] JILA, National Institute of Standards and Technology, and University of Colorado Department of Physics, University of Colorado, Boulder.
[2] IMRA America Inc., 1044 Woodridge Avenue, Ann Arbor, Michigan 48105, USA.
[3] Currently at Institute for Lasers, Life and Biophotonics, Vrije Universiteit Amsterdam, De Boelelaan 1081, 1081HV Amsterdam, The Netherlands.
* These authors have contributed equally to this work.

The existence of the XUV comb structure is critically dependent on the phase coherence of the HHG process. While the temporal coherence of HHG in an isolated pulse has been studied extensively[16,17], the requirements are more stringent for XUV comb generation where the phase coherence must be maintained over many consecutive pulses. This issue has been a topic of investigation since first demonstrations of intracavity HHG[3,4], where only the phase coherence of the third harmonic was verified. In subsequent work, pulse-to-pulse temporal coherence of the seventh harmonic at 152 nm, below the ionization threshold, was demonstrated interferometrically[18]. Recently, the phase coherence of above-threshold harmonics was utilized for two-pulse Ramsey spectroscopy of helium at 51 nm[19]. We note that this approach differs fundamentally from the work presented here. The two-pulse Ramsey sequence is a phase measurement and, as a result, is susceptible to systematic phase errors between the two pulses that lead to apparent frequency shifts. Thus, all sources of phase errors must be considered and characterised before an absolute frequency measurement can be made. In contrast, a frequency comb consisting of an infinite pulse train is an optical frequency synthesizer with equally spaced frequency comb teeth, defined by only two numbers: the repetition frequency ($f_{rep}$) of the pulse train and the carrier envelope offset frequency ($f_{ceo}$). Thus, the phase instabilities can lead to broadening of the comb teeth, but the absolute frequencies of the XUV comb remain simply and robustly determined. It is precisely this characteristic that led to the recent frequency comb revolution in metrology even though multi-pulse Ramsey spectroscopy in the visible and infrared spectral regions has existed since the late 1970's[20]. In this letter, we demonstrate for the first time the extension of frequency comb techniques to the XUV spectral region, which will bring about a new era in short-wavelength metrology.

In intracavity HHG, a high-average power infrared (IR) frequency comb[21] is coupled into a dispersion-controlled optical cavity (Fig. 1), which allows for passive coherent buildup of pulses at the full repetition rate (154 MHz). The high repetition rate of intracavity HHG should in principle provide increased average flux in comparison to conventional single pass HHG systems, which typically employ lasers with repetition frequencies and average powers that are smaller by orders of magnitude. However, until now, the presence of the optical cavity has presented challenges to matching the average flux of these systems due to optical damage problems associated with the large average intracavity power, as well as the nonlinear-response of the HHG medium. At high intensities and gas densities required to optimize the XUV flux, the plasma density at the intracavity focus reaches levels that can lead to various nonlinearities such as optical bistability, self phase modulation, and pulse distortion. Recently, these effects have been studied in detail with both simulations and experiments[22,23]. These effects clamp the achievable intracavity peak power and lead to instabilities in the lock between the comb and the cavity, limiting the attainable XUV flux and coherence time. Since all of these problems are magnified by the finesse of the cavity, a robust method for mitigating them is to decrease the finesse. However, this requires a more powerful laser to excite the enhancement cavity in order to keep the intracavity pulse energy high. This challenge has been answered by recent advances in high repetition rate, high average power chirped-pulse amplified $Yb^+$ fibre lasers[21,24-26].

Our ability to reach high intracavity pulse energies while minimizing the deleterious effects of a large plasma density has resulted in record-level power in high harmonics spectrally separated for spectroscopy. We measure the power in the 15$^{th}$ harmonic to be 21 μW with Xe as the target gas (Fig. 2), which is more than an order of magnitude improvement over previous intracavity HHG results[27,28]. Similar outcoupled power (not spectrally separated) was recently reported[29]. We have also observed shorter wavelength radiation by using Kr as the target gas for HHG. Due to the higher ionization potential of Kr, we were able to detect the 25$^{th}$ and the 27$^{th}$ harmonic orders with 8 kW of intracavity power (peak intensity of 9 x 10$^{13}$ W/cm$^2$) as shown in Fig. 2c, extending intracavity HHG to photon energies above 30 eV. As shown in Fig. 2d, there are only modest improvements in power for most harmonics as the intracavity IR power is increased above 5.5 kW. As the plasma density increases, depletion of the on-axis neutral atomic density and decrease in phase matching result in the XUV power clamp. These results show that future gains in harmonic power in a similar system will most likely be achieved by increasing the focal area at the interaction region.

To demonstrate the comb structure of XUV radiation, we have spectrally resolved the $3s^23p^6$ $J=0$ ➔ $3s^23p^55d$ $J=1$ electric dipole (E1) transition in argon with an upper state energy of 121,932.8 cm$^{-1}$ (~82 nm), and the $2s^22p^6$ $J=0$ ➔ $2s^22p^54s$ $J=1$ E1 transition in neon with an upper state energy of 159,534.6 cm$^{-1}$ (~63 nm) via resonance fluorescence spectroscopy. These transitions lie within the 13$^{th}$ and 17$^{th}$ harmonic bandwidth of our source, respectively. In order to reduce the Doppler width of the transition below the repetition frequency of the comb, we use pulsed supersonic atomic beams propagating orthogonal to the XUV beam. The average harmonic power available for spectroscopy is ~ 1 µW due to a single reflection from a mirror to steer the XUV beam into the interaction region and several pinholes to isolate the central part of the beam that contains the phase-coherent short trajectories[18]. Moreover, due to the electric dipole nature of these transitions, only a single comb tooth with ~10$^{-5}$ of the total power contributes to the signal. Nevertheless, we have achieved a signal to scattered photon background ratio of ~1 using a photomultiplier tube to directly detect the XUV spontaneous emission to the ground state.

The fundamental infrared comb is controlled by phase locking a comb tooth to a continuous wave neodymium-doped yttrium aluminium garnet (Nd:YAG) laser by feedback onto the cavity length, which mainly actuates on $f_{rep}$ of the comb. The Nd:YAG laser frequency ($\nu_{YAG}$) is referenced to a molecular iodine transition and serves as an absolute optical frequency marker with an uncertainty of 5 kHz[30]. The comb $f_{ceo}$ has a free running linewidth of 15 kHz[21] and is adjusted to optimize intracavity power, but is otherwise unlocked. Its value is deduced from the knowledge of $\nu_{YAG}$ and $f_{rep}$, which is measured against the NIST hydrogen maser. Figure 3 shows the fluorescence signal as the phase-locked-loop offset frequency, $\delta$, is slowly changed, and XUV comb teeth are scanned across the atomic resonance for two different collimation angles of the atomic beam. With the background subtracted, the signal contrast is 100%, confirming the sharp comb structure. Least squares fits for the argon data reveal a Gaussian full width at half maximum of 47±5 and 11±1 MHz, which are consistent with what would be expected from the geometric collimation. Thus, the observed linewidth places only an upper bound on the comb teeth linewidth of less than 10 MHz. The statistical uncertainty in the line centre determination for the narrower lineshape is 500 kHz, demonstrating an unprecedented fractional frequency precision of better than 2 × 10$^{-10}$ in the XUV.

To determine the absolute frequency of the argon transition, it is necessary to determine the comb tooth number in addition to $f_{rep}$ and $f_{ceo}$. This requires making several measurements at different values of $f_{rep}$ and keeping track of the change in the comb tooth number[31]. Fig. 4 documents a total of 10 measurements at different values of $f_{rep}$ chosen to provide absolute comb mode determination. The data analysis is carried out by noting that the fixed point of the infrared comb (comb tooth phase locked to the Nd:YAG laser) is multiplied up by the HHG process to create corresponding fixed points for every harmonic comb. These fixed points are offset from $q \times \nu_{YAG}$ by $\Delta = q \times \delta$ (modulo $f_{rep}$), where $q$ is the harmonic order. Thus, the argon transition frequency can be determined as:

$$\nu_{Ar} = n \times f_{rep} + 13 \times f_{ceo} = \Delta n \times f_{rep} + (13 \times \nu_{YAG} + \Delta),$$

where $n$ is the absolute comb tooth number and $\Delta n$ is the comb tooth number difference between the 13$^{th}$ harmonic fixed point and the actual comb tooth probing the argon transition. This point of view not only makes the analysis immune to drifts in $f_{ceo}$, but also highlights the direct frequency link between the infrared and the higher harmonic combs.

Once $\Delta n$ is determined, we correct the final value for residual Doppler shifts arising from imperfect orthogonal alignment of the laser and the atomic beam. This is accomplished by comparing the measured transition frequency of a pure argon beam (velocity of 520±20 m/s) with that of a helium argon mixture (5:1 He-Ar partial pressures, velocity 740±60 m/s) and translating the final atomic beam collimation slit until the two agree, ensuring orthogonality between the laser and the atomic beams. Due to the small fractional change in the velocity of the two beams, the uncertainty in the procedure is ~ 3 MHz. All other systematic shifts such as recoil, density, and Stark shifts, as well as the contribution due to the uncertainty in $\nu_{YAG}$, are estimated to be smaller than 1 MHz. We have also binned our data for different laser operation conditions and find no systematic shifts at the 2 MHz level for 30% change in xenon backing pressure or 20% change in the intracavity power for HHG, again highlighting the robustness of the XUV comb for frequency measurements. The final absolute frequency for this transition is 3,655,454,073±3 MHz, which agrees well with previous measurements that have an uncertainty of 2.3 GHz[32].

In conclusion, we have generated >200 μW of average power per harmonic via HHG in a femtosecond enhancement cavity. This has enabled the verification of the comb structure of the XUV light via direct frequency comb spectroscopy. We have also measured the absolute transition frequency by making multiple measurements at different repetition frequencies, demonstrating that the XUV frequency comb has become a mature and robust tool. The upper bound on the 13$^{th}$ harmonic comb linewidth is less than 10 MHz, limited by the atomic thermal motion. In the future, the heterodyne beat signals between two intracavity HHG sources will be used to overcome this limitation, improving the phase noise characterization by orders of magnitude. With these developments, ultrahigh precision spectroscopy in the XUV is firmly within grasp, enabling a wide range of applications.

**Methods Summary**

The Yb$^+$ fibre comb delivers 120 fs pulses at 154 MHz repetition frequency with a maximum average power of 80 W[21]. For the studies presented here, we run the laser at ~ 30 W. The enhancement cavity finesse is ≈ 400. Target gas is injected at the intracavity focus with a glass nozzle with a 100 μm aperture and a backing pressure of ~2 atm. The XUV light is coupled out of the cavity with an intracavity diffraction grating[27] with 10% efficiency.

The 13$^{th}$ harmonic power is measured with an XUV photodiode (IRD Inc. AXUV100In/MgF$_2$), which was confirmed to be insensitive to scattered light in the present geometry. The power in the other harmonics is determined using a plate coated with sodium salicylate, which fluoresces at 420 nm when excited by XUV radiation with uniform quantum efficiency between 40 and 100 nm[33]. The fluorescence is imaged using a CCD camera and calibrated using the 13$^{th}$ harmonic, which can be measured with either the photodiode or the CCD. The estimated calibration uncertainty is 20-30%.

The supersonic argon source consists of a pulsed valve nozzle (0.5mm diameter), a 0.5 mm diameter skimmer, and a secondary slit located 33 cm from the skimmer. Both the slit width and position are adjustable in order to study the Doppler width and shifts. The valve generates 500 μs long pulses at a repetition rate of 20 Hz with an argon backing pressure of 0.5 atm.

The fluorescence detector consists of a lightpipe coated with sodium salicylate, photomultiplier tube optimized for single photon detection (Hamamatsu H11123) and a pulse counter (SR400). The pulse counter has two gates: one coincident with the Ar pulse arrival time to measure signal, and another between Ar pulses to measure background. Most of the laser light traverses unimpeded through the interaction region and is detected with a second XUV photodiode, which provides the normalization signal. Individual frequency scans take ~150 s. Results presented in Fig. 3 are averages of 3 – 10 scans for a total acquisition time of 8 to 25 minutes.

**References**

1. Udem, T., Holzwarth, R. & Hänsch, T.W. Optical frequency metrology. *Nature* **416**, 233-237 (2002).
2. Cundiff, S.T. & Ye, J. Colloquium: Femtosecond optical frequency combs. *Rev. Mod. Phys.* **75**, 325 (2003).
3. Jones, R.J., Moll, K.D., Thorpe, M.J. & Ye, J. Phase-Coherent Frequency Combs in the Vacuum Ultraviolet via High-Harmonic Generation inside a Femtosecond Enhancement Cavity. *Phys. Rev. Lett.* **94**, 193201 (2005).
4. Gohle, C. et al. A frequency comb in the extreme ultraviolet. *Nature* **436**, 234-237 (2005).
5. Merkt, F. and Softley, T. P. Final-State Interactions In The Zero-Kinetic-Energy-Photoelectron Spectrum Of H$_2$. *J. Chem. Phys.* **96**, 4149-4156 (1992).
6. Herrmann, M. et al. Feasibility of coherent XUV spectroscopy on the 1S-2S transition in singly ionized helium. *Phys. Rev. A* **79**, 052505 (2009).
7. Eyler, E.E. et al. Prospects for precision measurements of atomic helium using direct frequency comb spectroscopy. *Eur. Phys. J. D* **48**, 43-45 (2008).
8. Peik, E. & Tamm, C. Nuclear laser spectroscopy of the 3.5 eV transition in Th-229. *Europhys. Lett.* **61**, 181-186 (2003).
9. Rellergert, W.G. et al. Constraining the Evolution of the Fundamental Constants with a Solid-State Optical Frequency Reference Based on the $^{229}$Th Nucleus. *Phys. Rev. Lett.* **104**, 200802 (2010).
10. Campbell, C.J., Radnaev, A.G. & Kuzmich, A. Wigner Crystals of $^{229}$Th for Optical Excitation of the Nuclear Isomer. *Phys. Rev. Lett.* **106**, 223001 (2011).
11. Murphy, M.T., Webb, J.K. & Flambaum, V.V. Further evidence for a variable fine-structure constant from Keck/HIRES QSO absorption spectra. *Monthly Notices of the Royal Astronomical Society* **345**, 609-638 (2003).
12. Webb, J.K. et al. Evidence for spatial variation of the fine structure constant. submitted to *Phys. Rev. Lett.* http://arxiv.org/abs/1008.3907 (2010).
13. J. C. Berengut, V. A. Dzuba, V.V. Flambaum. Enhanced laboratory sensitivity to variation of the fine-structure

constant using highly charged ion. *Phys. Rev. Lett.* **105**, 120801 (2010).
14. J. C. Berengut, V. A. Dzuba, V.V. Flambaum, A. Ong. Electron-hole transitions in multiply charged ions for precision laser spectroscopy and searching for variations in alpha. *Phys. Rev. Lett.* **106**, 210802 (2011).
15. Krausz, F. & Ivanov, M. Attosecond physics. *Rev. Mod. Phys.* **81**, 163 (2009).
16. Bellini, M. et al. Temporal Coherence of Ultrashort High-Order Harmonic Pulses. *Phys. Rev. Lett.* **81**, 297-300 (1998).
17. Mairesse, Y. et al. Attosecond Synchronization of High-Harmonic Soft X-rays. *Science* **302**, 1540 -1543 (2003).
18. Yost, D.C. et al. Vacuum-ultraviolet frequency combs from below-threshold harmonics. *Nat Phys* **5**, 815-820 (2009).
19. Kandula, D.Z., Gohle, C., Pinkert, T.J., Ubachs, W. & Eikema, K.S.E. Extreme ultraviolet frequency comb metrology. *Phy. Rev. Lett.* **105**, 063001 (2010).
20. Eckstein, J.N., Ferguson, A.I. & Hänsch, T.W. High-Resolution Two-Photon Spectroscopy with Picosecond Light Pulses. *Phys. Rev. Lett.* **40**, 847 (1978).
21. Ruehl, A., Marcinkevicius, A., Fermann, M.E. & Hartl, I. 80 W, 120 fs Yb-fiber frequency comb. *Opt. Lett.* **35**, 3015-3017 (2010).
22. Allison T. K. , Cingöz A., Yost D.C., Ye J., Cavity extreme nonlinear optics. submitted to *Phys. Rev. Lett.* http://arxiv.org/abs/1105.4195 (2011).
23. Carlson, D.R., Lee, J., Mongelli, J., Wright, E.M. & Jones, R.J. Intracavity ionization and pulse formation in femtosecond enhancement cavities. *Opt. Lett.* **36**, 2991-2993 (2011).
24. Hartl, I. et al. Cavity-enhanced similariton Yb-fiber laser frequency comb: $3 \times 10^{14}$ W/cm$^2$ peak intensity at 136 MHz. *Opt. Lett.* **32**, 2870-2872 (2007).
25. Schibli T. R. et al. Optical frequency comb with submillihertz linewidth and more than 10 W average power. *Nat Photon* **2**, 355-359 (2008).
26. Eidam, T. et al. Femtosecond fiber CPA system emitting 830 W average output power. *Opt. Lett.* **35**, 94-96 (2010).
27. Yost, D.C., Schibli, T.R. & Ye, J. Efficient output coupling of intracavity high-harmonic generation. *Opt. Lett.* **33**, 1099-1101 (2008).
28. Ozawa, A. et al. High Harmonic Frequency Combs for High Resolution Spectroscopy. *Phys. Rev. Lett.* **100**, 253901 (2008).
29. Jones, J. Intracavity high harmonic generation with fs frequency combs. In *High Intensity Lasers and High Field Phenomena*, OSA Technical Digest (CD) (Optical Society of America, 2011), paper HFB5.
30. Ye, J., Ma, L.-S., Hall, J. L. Molecular Iodine Clock. Phys. Rev. Lett. **87**, 270801 (2001).
31. Stowe M. *Direct Frequency Comb Spectroscopy and High-Resolution Coherent Control*. Ph.D. Thesis. University of Colorado, Boulder (2008).
32. Minnhagen, L. Spectrum and the energy levels of neutral argon, Ar I. *J. Opt. Soc. Am.* **63**, 1185-1198 (1973).
33. Samson, J. A. R. *Techniques of Vacuum Ultraviolet Spectroscopy*. (VUV Associates, 1990).

**Acknowledgments:** This research is funded by the DARPA, AFOSR, NIST, and NSF. A. Cingöz and T. K. Allison are National Research Council postdoctoral fellows. A. Ruehl acknowledges funding from the Alexander von Humboldt Foundation (Germany).

**Author Contributions:** A.C., D.C.Y., T.K.A., and J.Y. conceived of, designed, and carried out the XUV power and spectroscopy measurements. A.R., M.E.F., and I.H. designed and built the Yb$^+$ fibre laser. All authors discussed the results and commented on the manuscript.

**Author Information:** Reprints and permissions information is available at www.nature.com/reprints. The authors have no competing financial interest. Any mention of commercial products does not constitute an endorsement by NIST. Correspondence should be addressed to A.C. (acingoz@jila.colorado.edu) or J.Y. (junye@jila.colorado.edu).

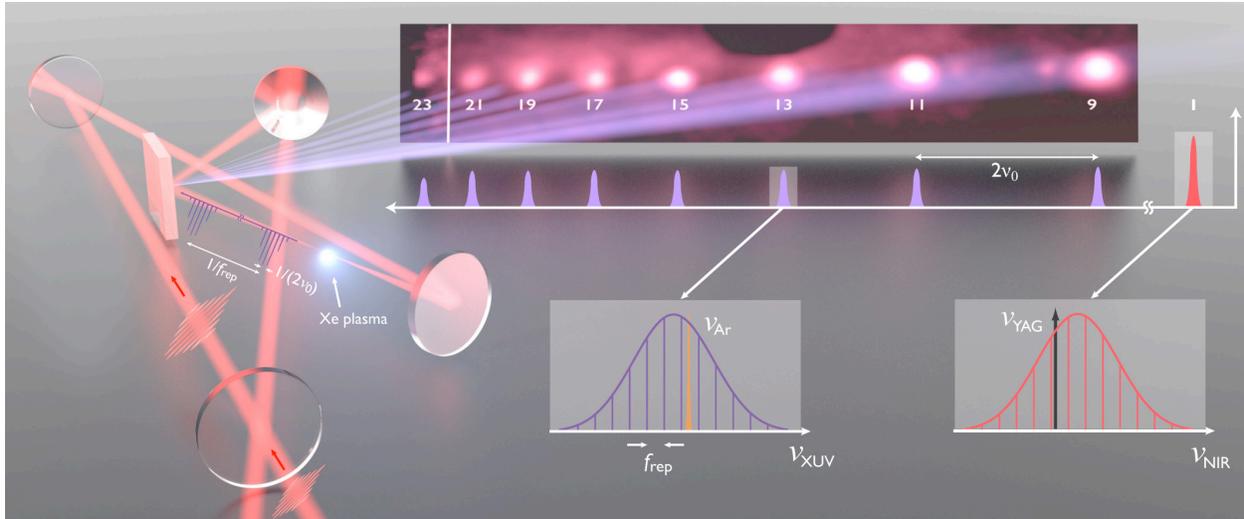

**Figure 1: Intracavity High Harmonic Generation.** An infrared frequency comb is passively amplified in a power buildup cavity with a unique mirror-diffraction grating hybrid output coupler for the XUV[27]. Xenon gas is introduced at the tight focus of the cavity. Also shown is the resultant HHG spectrum from a single infrared pulse, which consists of odd harmonics of the infrared carrier frequency $\nu_0$. By using an infinite train of pulses, we create a new comb structure in every harmonic at the repetition frequency ($f_{rep}$) of the pulse train. This comb structure is stabilized using a continuous wave laser ($\nu_{YAG}$) and slowly scanned over an argon transition ($\nu_{Ar}$) at 82 nm.

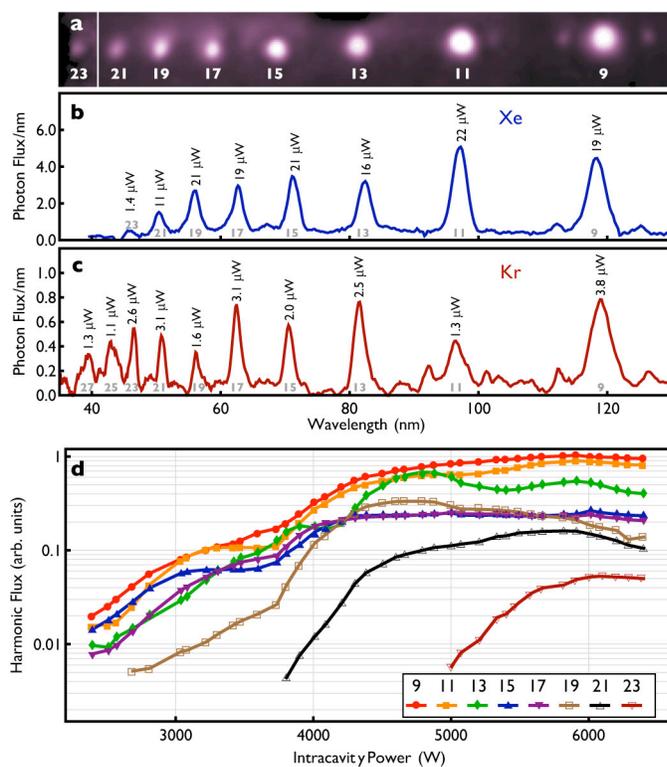

**Figure 2: Power scaling results: a.** A characteristic image of the harmonics in Xe dispersed on a plate coated with fluorescent material. The threshold of the image was adjusted for the 23$^{rd}$ harmonic to make it more visible. A lineout of a CCD image for **b.** Xe and **c.** Kr target gas with corresponding integrated harmonic powers. The observed widths of the harmonic orders are given by the instrument wavelength resolution. **d.** Harmonic powers as a function of infrared intracavity power showing a marginal increase above 5.5 kW. The oscillations visible for some harmonics are due to quantum path interference[18].

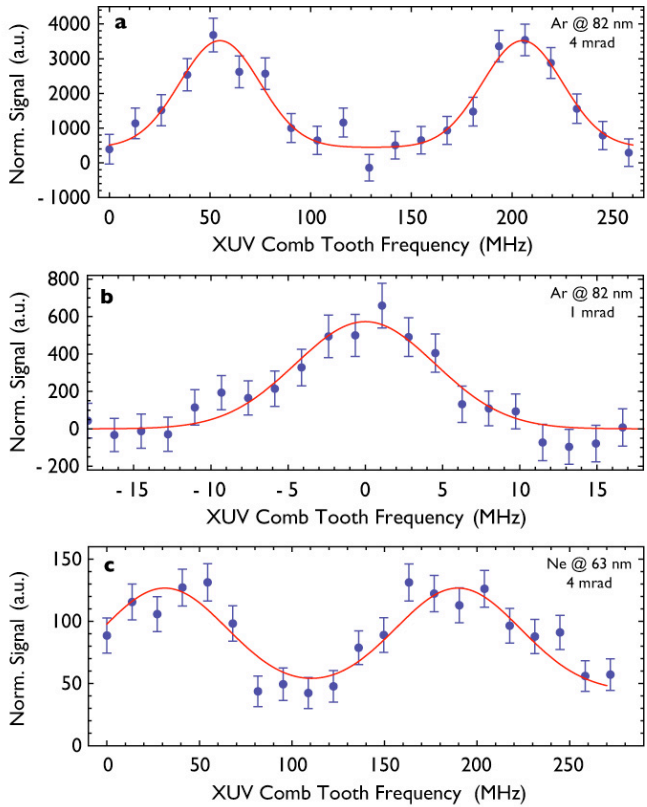

**Figure 3: Atomic fluorescence signal:** The photomultiplier signal as a function of an XUV comb tooth frequency for **a.** ~4 mrad and **b.** ~1 mrad argon beam collimation angle. The photomultiplier signal is normalized to the incident XUV power and the background signal subtracted (see Methods section). The error bars represent the shot noise due to fluorescence photon counting statistics. The red curves are least square fits to a Gaussian model. **c.** the signal for the neon transition with ~4 mrad collimation angle. The contrast and signal-to-noise are lower due to the increased Doppler width and the smaller absorption cross-section, respectively.

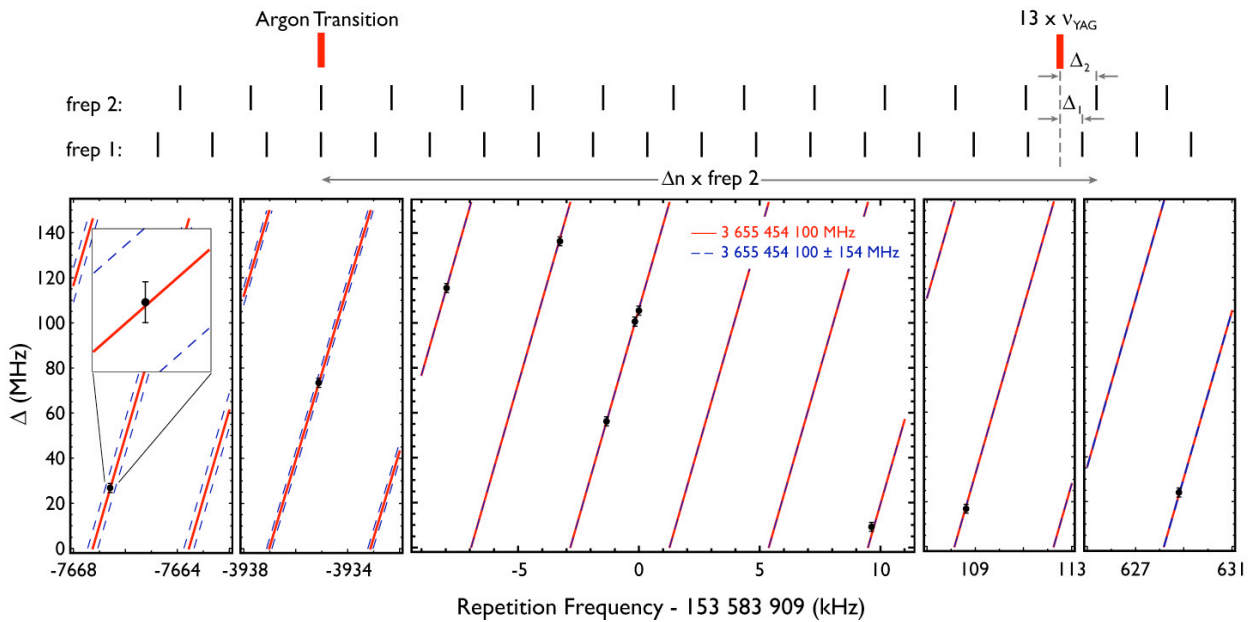

**Figure 4: Absolute frequency determination:** Measurement of the resonance centre frequency at 10 different values of $f_{rep}$ parameterized by the offset, $\Delta$, of the 13<sup>th</sup> harmonic fixed point from $13 \times \nu_{YAG}$. Also plotted are three lines: a solid red line showing the expected location of the resonance centers for the extracted comb tooth number, $n$, and two blue dashed lines showing the expected locations for $n\pm1$. The **inset** shows a zoomed in version of a measurement, which demonstrates that the comb tooth number is unambiguously determined.